# STM Single Atom/Molecule Manipulation and Its Application to Nanoscience and Technology

Saw-Wai Hla

*Department of Physics and Astronomy, Nanoscale and Quantum Phenomena Institute, Ohio University, Athens, OH-45701, USA; email: hla@helios.phy.ohiou.edu*

**Abstract:**

Single atom/molecule manipulation with a scanning-tunneling-microscope (STM) tip is an innovative experimental technique of nanoscience. Using STM-tip as an engineering or analytical tool, artificial atomic-scale structures can be fabricated, novel quantum phenomena can be probed, and properties of single atoms and molecules can be studied at an atomic level. The STM manipulations can be performed by precisely controlling tip-sample interactions, by using tunneling electrons, or electric field between the tip and sample. In this article, various STM manipulation techniques and some of their applications based on the author's experience are described, and the impact of this research area on nanoscience and technology is discussed.

**\* Correspondence to;** Saw-Wai Hla, Physics and Astronomy Dept., Ohio University, Clippinger 252 C, Athens, OH-45701, USA. Tel: 740 593 1727, Fax: 740 593 0433.

## I. INTRODUCTION

In the mid 20th century, the possibility to image or to *see* an atom was a matter of great debate. Only after the Nobel-award winning invention of Scanning Tunneling Microscope (STM) by Binnig and Rohrer in the early 1980s, atomic landscapes of material surfaces could be imaged in real space[9]. The operation principle of STM is based on a quantum mechanical phenomenon known as "tunneling"[13]. When a sharp needle (tip) is placed less than 1 nm distance from a conducting material surface (sample) and a voltage is applied between them, the electrons can tunnel between the tip and the sample through the narrow vacuum barrier. Since the tunneling current exponentially varies with the tip-sample distance, a tiny change in the distance less than a fraction of the atomic length can be detected[13]. During STM imaging, location of STM-tip at the proximity of the surface often causes perturbations due to tip-sample interactions. In a normal imaging mode, these perturbations are not desirable. Since the beginning of 1990s, these undesired perturbations became one of the most fascinating subjects to be pursued by scientists[22]: manipulation of atoms and molecules on surfaces. Using STM manipulation techniques, quantum structures can be constructed on an atom-by-atom basis[12, 17, 34-37, 61], single molecules can be synthesized on a one-molecule-at-a-time basis[31, 39, 40, 47, 57], and detailed physical/chemical properties of atoms/molecules, which are elusive to other experimental measurements, can be accessed at an atomic level[3, 11, 18, 23, 27, 29, 32, 42, 48, 52, 55, 72-78]. Now a day, STM is an instrument not only used to *see* individual atoms by imaging, but also used to *touch* and *take* the atoms or to *hear* their vibration by means of manipulation. In this perspective, STM can be considered as the *eyes, hands* and *ears* of the scientists connecting our macroscopic world to the exciting atomic and nanoscopic world.

## II. EXPERIMENTAL SET-UP

Manipulation of single atoms and molecules with a STM-tip generally requires atomically clean surfaces and an extreme stability at the tip-sample junction. Therefore, ultra-high-vacuum (UHV) environments and low substrate temperatures are favored in most manipulation experiments. Fig. 1 illustrates a custom-built low temperature STM system capable of performing a variety of manipulation procedures described in this article. This LT-STM system, which is built based on the design of Gerhard Meyer[66], includes a Bescoke-Beetle type STM scanner[7] attached to the base part of a liquid-helium bath cryostat. The system is equipped with a state-of-the-art UHV facilities, sample cleaning and atom/molecule deposition sources. The STM scanner is housed inside two thermal radiation shields, which also act as cryogenic pumps providing a local pressure well below 4 x 10^-11 Torr. This allows maintaining atomically clean sample condition for a long period of time. Electrochemically etched polycrystalline tungsten wires are used as the tips in most experiments described here. The tip-apex is usually prepared by using in-situ tip formation procedure[36] described in the next section.

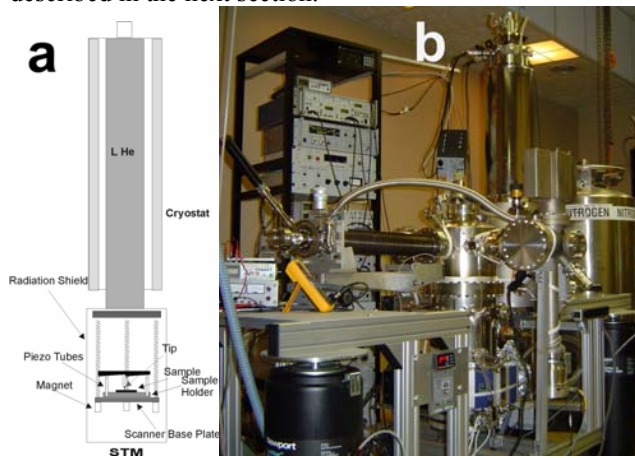

**Fig. 1.** Low-temperature STM for single atom/molecule manipulation. (a) A schematic drawing of STM scanner set-up. (b) An image showing a custom-built UHV-LT-STM system at the author's lab.

## III. PREPARATION FOR MANIPULATION BY TIP-CRASH

When an STM-tip plunges into a material surface, which is often known as "tip-crash", it causes a crater on the surface. Normally, a tip-crash is an undesired event because it can destroy the sample surface. However, the STM tip-crash procedure is useful in several applications including nanoindentation to test hardness of material surfaces[14, 24, 26, 46, 54, 90], and nanowire formation between the tip and sample[15, 50, 88]. Recently, controlled tip-crash procedures are demonstrated to be useful in reshaping the tip-apex and extracting individual atoms from the native substrate[36].

The shape, structure and chemical identity of the tip-apex play a vital role in STM manipulations. Therefore, formation of an atomically sharp tip-apex with known chemical identity is a necessary step for a manipulation experiment. To form the tip-apex, the tip is dipped into a metallic substrate for a few nm under a high bias (>4 V) condition. The thermal energy produced by the tip-crash causes reshaping of the tip-apex into a sharper profile. During the process, the tip-apex is also coated with the substrate material and hence, its chemical identity can be known (Fig. 2). For extraction of individual atoms from the native substrate, the controlled tip-crash is performed under a low tunneling bias condition. After the procedure, scattered individual atoms can be observed near the tip-surface contact area[36] (Fig. 3). In case of a tip-crash using a silver coated tungsten-tip on a Ag(111) surface, mostly silver atoms are extracted[36]. These atoms can be used as the basic building blocks of atomistic constructions. If the single atoms can be locally extracted from the native substrate, additional atom deposition is no longer necessary and hence it has an advantage in some nanoscale experiments.

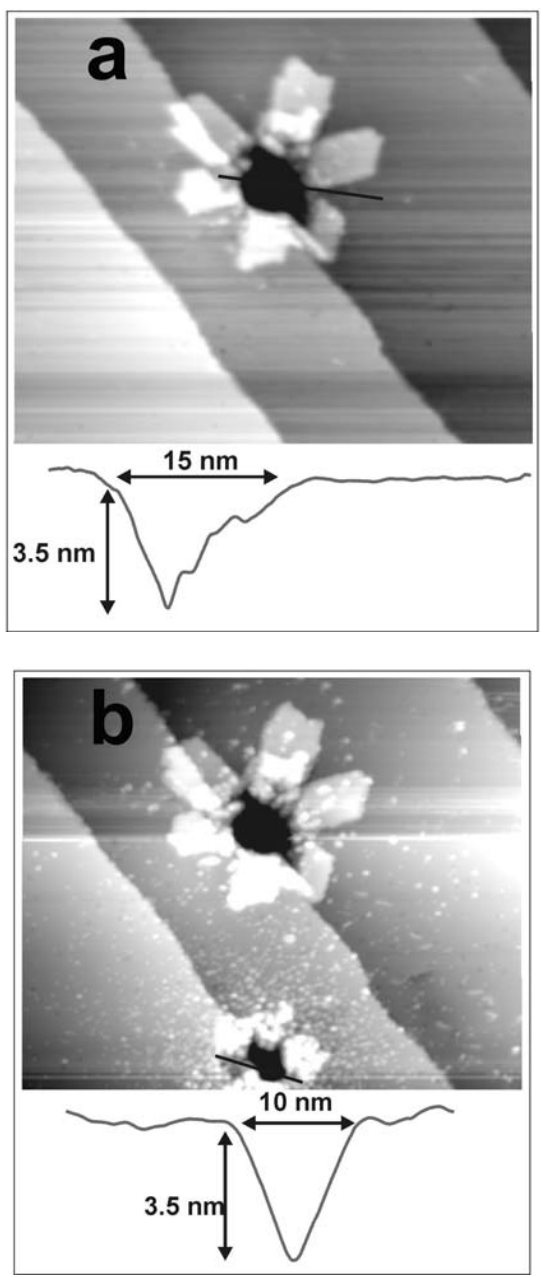

**Fig. 2** STM images of tip indentations. (a) An STM image of Ag(111) surface after a tip-crash. (b) An STM image acquired after the second tip-crash, which produces a sharper triangular shape hole. [Imaging parameters: $V_t$ = 0.29 V, $I_t$ = 1.7 nA, scan area = 120 x 120 nm$^2$].

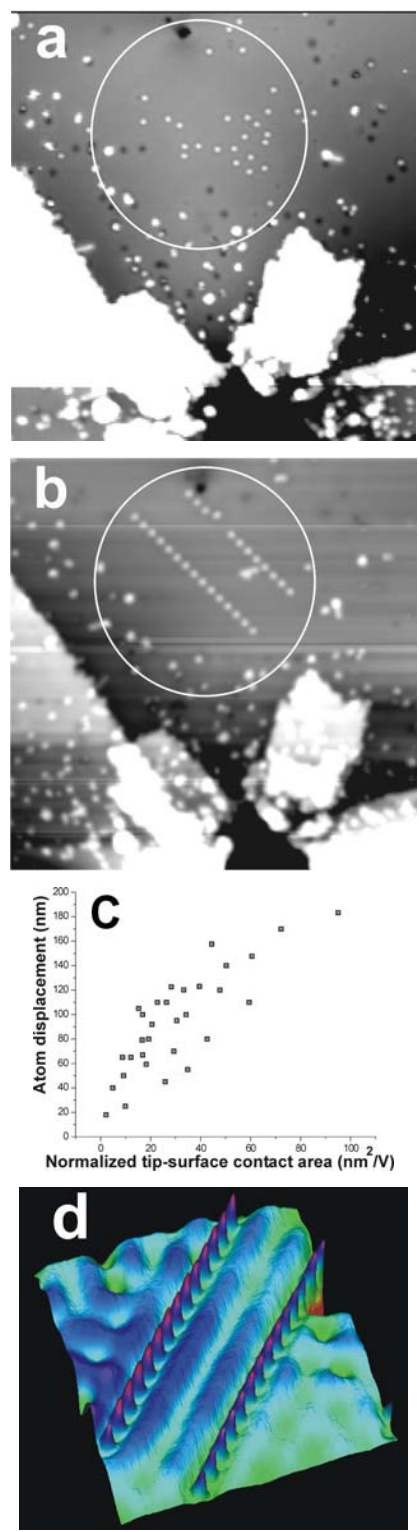

**Fig. 3** (a) An STM image shows scattered silver atoms produced by a tip-crash. (b) The atoms are repositioned with the STM-tip to construct a quantum structure. (c) The normalized tip-sample contact area vs. atom displacement plot. (d) A three dimensional representation of the constructed structure.

## IV. LATERAL MANIPULATION

An STM manipulation procedure to relocate single atoms/molecules across a surface is known as the "lateral manipulation" (LM) [5, 8, 10, 11, 22, 30, 35, 37, 41, 43, 45, 51, 53, 58, 59, 61, 63-68, 70, 71, 84-86] (Fig. 4). The first example of LM was demonstrated by Eigler and Schweizer in 1990 by writing the "IBM" company logo with Xe atoms on a Ni(110) surface[22]. An extremely fine control over the tip-atom-surface interaction is necessary to achieve the atomic-scale precision. A typical LM procedure involves three steps: 1) vertically approaching the tip towards the manipulated atom to increase the tip-atom interaction, 2) scanning the tip parallel to the surface where the atom moves under the influence of the tip, and 3) retracting the tip back to the normal image-height thereby the atom is left at the final location on the surface (Fig. 4).

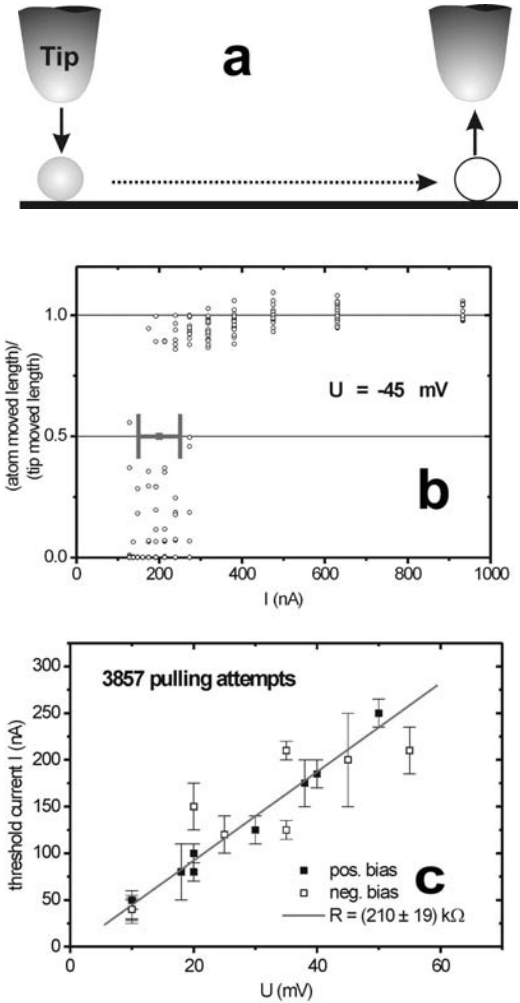

**Fig. 4** Lateral manipulation (LM). (a) A schematic drawing shows the tip action during LM. (b) The probability to move an atom vs. tunneling current at -45 mV. (c) At the small bias range, the threshold current linearly varies with the bias. $R_t$ can be determined from the slope of this curve.

In a LM process, the tunneling resistance ($R_t$) is used to indicate the tip-atom distance, and to qualitatively estimate the tip-atom interaction strength[5, 37]. Large and small $R_t$ correspond to far and close tip-atom distances, or strong and weak tip-atom interactions, respectively. The minimum tip-atom distance required for a manipulation can be determined by measuring a threshold tunneling current ($I_T$) to move an atom at a fixed bias (Fig.4b)[37]. Fig. 4c shows a plot of $I_T$ as a function of applied bias. Here, each data point is determined by plotting a curve, like the one shown in Fig.4b. From the slope of this curve, a threshold tunneling resistance required to move the atom can be determined. In case of a silver atom manipulation on Ag(111), $R_t = 184 \pm 8$ k$\Omega$ is necessary[37]. This $R_t$ corresponds to a distance of 1.9 Å between the edges of van-der-Waals radii of tip-apex and manipulated atom. Since the atomic orbitals of tip-apex and manipulated atoms are overlapping at this distance, a weak chemical bond is formed. The attractive force used in the "pulling" manipulation is originated from this chemical nature of interaction.

The nature of atom movements and the type of tip-atom interactions during a LM process can be determined from the STM feedback or tunneling current signals[5, 10, 11, 35, 37, 41, 51]. The nature of forces involved in a LM process and the mechanisms of atom movement have been analyzed by the group of Rieder[5]. Three basic LM modes, "pushing", "pulling" and "sliding", has been distinguished[5]. In the "pulling" mode, the atom follows the tip due to an attractive tip-atom interaction. In the "pushing" mode, a repulsive tip-atom interaction drives the atom to move in front of the tip. In the "sliding" mode, the atom is virtually bound to or trapped under the tip and it moves smoothly across the surface together with the tip.

A schematic diagram of force components involved in an atom manipulation process is illustrated in Fig. 5a. Here, the STM is set in the constant current scanning mode. The tip is initially located directly above the atom and hence, only a *vertical force component* ($F_\perp$) exists (Fig. 5a). The tip-atom distance is carefully chosen so that the attractive interaction between them is not strong enough to overcome the desorption barrier of the atom. Since the *lateral force component* ($F_\parallel$) is negligible at this point the atom will neither transfer to the tip nor move to the next adsorption site (ad-site) on the surface. When the tip moves parallel to the surface it passes over the manipulated atom, thereby tracing a part of the atomic contour. This action increases $F_\parallel$ and decreases $F_\perp$ (Fig. 5a). When $F_\parallel$ overcomes the hopping barrier of the atom, the atom hops to the next ad-site under the tip. This action alerts the STM feedback system to retract the tip in order to maintain the current constant causing an abrupt increase in the tip-height.

Repeating the sequence provides a saw-tooth like tip-height curve (Fig. 5b).

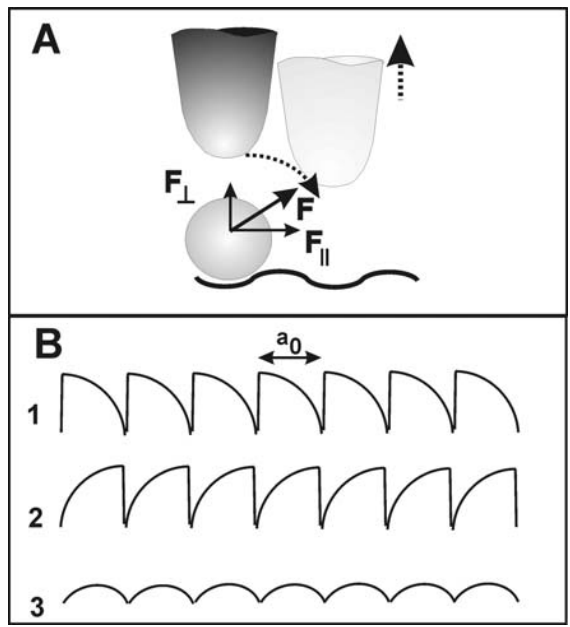

**Fig. 5.** (a) The drawing demonstrates the vertical and parallel force components involved in LM. (b) STM tip-height manipulation curves correspond to (1) pulling, (2) pushing, and (c) sliding modes.

In the "pushing" mode, the tip first climbs up the contour of the atom, and at some point the atom moves away from the tip due to the repulsion. As a result, an abrupt decrease in the tip-height occurs and repeating of this sequence provides another saw-tooth like signal with a reverse form of the "pulling" curve (Fig. 5b). The "pulling" and "pushing" manipulation signals can be recorded in the constant height scanning mode as well[51]. In that case the manipulation signal corresponds to the tunneling current intensity, not the tip-height variation. In both cases, the atom/molecule moves in a discontinuous manner by hopping the surface lattice sites[5, 37, 41]. The tip-height curves described above are also useful in determining the atom hoping distance during LM. For instance, the distance between two consecutive saw-tooth signals is a direct measure of the atom hopping distance.

The LM modes described above are mostly observed when an atom is manipulated along a close-packed row direction (a [110] direction) on fcc (111) surfaces. Atoms usually prefer to travel along close-packed directions on these surfaces due to a lower diffusion barrier as compared to other surface directions[51]. Sophisticated atom movement mechanisms occur when the atom is manipulated along the directions away from the surface close-packed directions[37, 51]. A silver atom movement along various manipulation directions on a Ag(111) surface is illustrated in Fig. 6[37]. For simplicity, we will describe the manipulation directions by means of deviation angle "θ" from a surface close-packed direction. Due to the Ag(111) surface symmetry the manipulation paths between θ = 0° and 30°, i.e. between the [110] and [211] surface directions (Fig. 6), can reveal the atom movements along any directions on this surface[37].

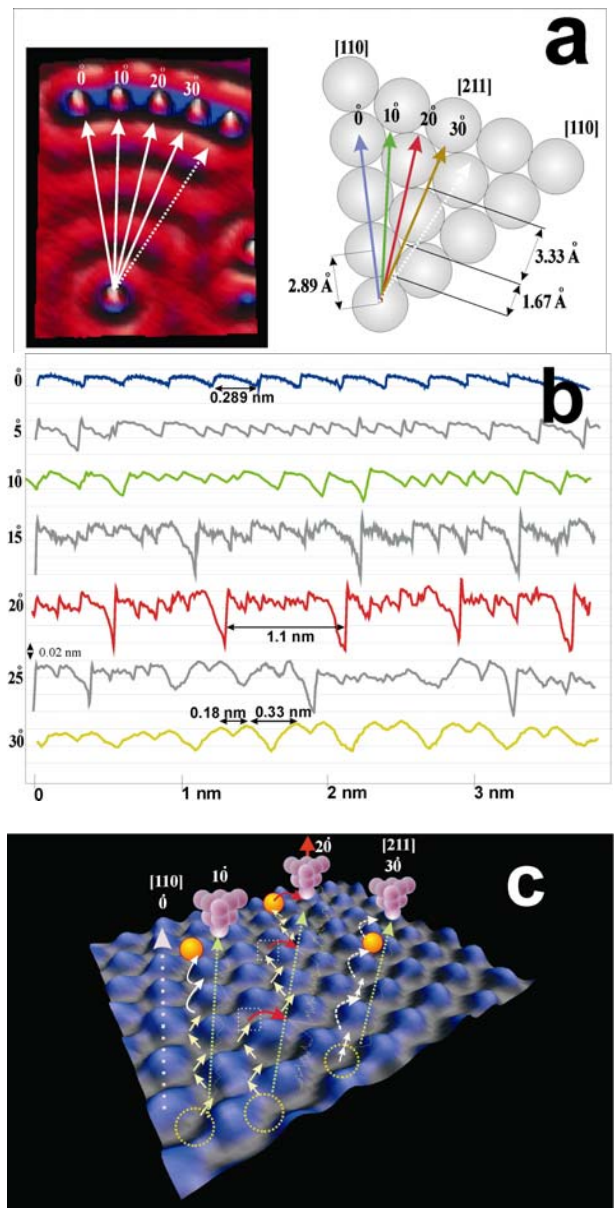

**Fig. 6.** (a) An STM image and a model demonstrate the tip-paths and Ag(111) surface geometry. (b) Atom manipulation signals along the tip-paths correspond to θ = 0°, 5°, 10°, 15°, 20°, 25°, and 30°. (c) The atom movements are demonstrated over an atomically resolved 3-D STM image of Ag(111). At θ = 5° and 10°, the small steps in manipulation signals are due to fcc-hcp site hopping. At θ = 15° and 20°, the periodic downward slopes followed by a tip-height increase is due to the "rest" and "jump" action of the atom. At 30°, the atom slides the first two near-by sites producing two consecutive bumps, and then it travels through a semi-cycle path to
5

reach the third site. At θ = 25°, the manipulation signal reveals both structures of θ = 20° and 30°.

At θ = 0°, the tip-path is along a surface close-packed direction and a typical "pulling" signal with single lattice-site hops of atom is observed (Fig. 6b). In θ = 5° and 10° tip-paths, the atom hops hcp and fcc lattice sites upon following the tip producing smaller hopping steps in the manipulation signal (Fig. 6b and 6c). The fcc-hcp hoping is induced due to a slight lateral displacement of the tip-apex either to the left or right side of the close-packed row where the atom is traveling. In θ = 15° and 20° tip-paths, the atom initially follows the tip by preferentially traveling along a surface close-packed direction. However, the atom can not travel along this path for a longer distance because the tip is moving into other direction away from the atom. At one point, the atom rests for a relatively longer time then the normal hopping time. This is due to a competition between the atom's preferential movement along the close-packed direction, and its tendency to follow the tip due to the attractive interaction. As the tip scans along the downward slope of the atom, the tip-height is reduced resulting in a deeper slope in the manipulation curve (Fig. 6b and 6c). This action of the tip increases the lateral force component, $F_∥$. When $F_∥$ overcomes the hopping barrier, the atom jumps to the adjacent surface close-packed row to follow the tip, which produces an abrupt increase in the tip-height. In this tip-path, the atom moves by repeating the "move-rest-jump" sequence in an approximately "zigzag" path to follow the tip[37, 51]. In θ =30° tip-path, the atom smoothly slides the first two neighboring hollow sites located 1.67 Å apart along a [211] direction (Fig. 6a and 6c). The third site is located relatively far distance (3.33 Å) then the nearest-neighbor distance (2.89 Å) of Ag(111). Instead of traveling directly to the third site, the atom travels approximately a semi-circle path by visiting two adjacent hcp-fcc hollow sites. Thus, all the atom visiting sites in this case are 1.67 Å distance apart to each other. Theoretical simulation reveals that these sites are energetically favored for the atom to move[51].

The atom manipulations described above correspond to one dimensional tip-paths. If LM is performed for a two-dimensional surface area, a series of manipulation signals (either tip-height or current signals) can be shown as an image[84]. For manipulation of large molecules, more complex movement behaviors can be observed[11, 35, 68, 70]. Depending on the molecule type and nature of intra-molecular bonding, large molecules can have more internal degrees of freedom. This allows large molecules to alter their molecular conformation to adapt the surface potential landscape changes during manipulation. The LM process can be used to probe detailed internal conformation changes of large molecules during their movement across a surface [11, 35, 68, 70].

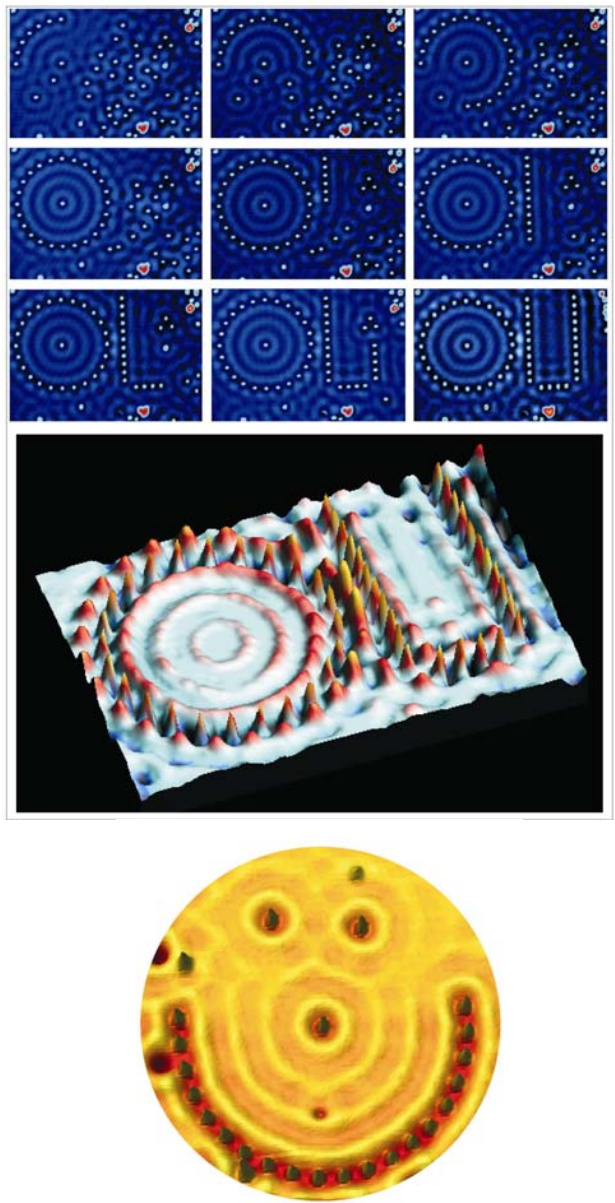

**Fig. 7.** OU logo writing sequence using individual silver atoms on a Ag(111) surface at 6 K (upper) and a three dimensional representation (middle) (42 nm x 26 nm area, 51 silver atoms are used). The "atomic smiley" image is written by using silver atoms on a Ag(111) surface at 5 K (32 nm diameter).

Manipulation of single atoms/molecules on surfaces allows construction of various quantum structures and investigation of novel phenomena associated to these structures[12, 17, 33, 35, 37, 61]. By constructing quantum corrals, the standing-electron-waves can be systematically studied[17], electron life-time inside an artificial structure can be measured[12], and quantum transport phenomenon can be probed[61]. By manipulating single CO molecules, Heinrich et al have constructed a molecular computing circuit that can perform logic operations[33]. The atom/molecule manipulation using the STM-tip is not only a scientific research but also applicable as an art. To exemplify the variety and novelty of atomistic constructions, an STM manipulation sequence of Ohio University logo (OU), and an "atomic smiley" created by the author are shown in Fig. 7.

## V. VERTICAL MANIPULATION

Vertical manipulation (VM) is another useful manipulation technique and it involves the transfer of atoms/molecules between the tip and surface[4, 6, 19, 21, 60, 65, 79, 87] (Fig. 8a). This process is closely related to desorption and subsequent adsorption of atoms/molecules on surfaces. An "atomic switch" realized by repeated transfer of a "Xe" atom between the STM tip and a Ni (110) substrate is the first example of VM[21]. The atom/molecule transfer process can be realized by using the electric field between the tip and sample, or by exciting with inelastic tunneling electrons, or by making tip-atom/molecule mechanical contact. The atom/molecule transfer mechanism can be explained by using a double potential well model (Fig. 8b)[79, 89]. At an imaging distance - at which the tip is roughly 6 Å above the surface - the manipulated atom/molecule has two possible stable positions, one at the surface and one at the tip-apex. Each position is represented by a potential well and the two potential wells are separated by a barrier in between (Fig. 8b). If the tip is positioned above the atom and an electric field is applied, then the shape of the double potential well changes to the one shown with the dashed line in Fig. 8b. In this case, the barrier between the two wells is reduced and the potential well at the tip-apex has a much lower energy. Now, the atom/molecule can easily transfer to the tip. By reversing the bias polarity, the minimum potential well can be shifted to the surface side. Then the dashed curve will reverse 180° from right to the left, and the atom/molecule can be transferred back to the surface. In case of VM performed by a tip-atom/molecule mechanical contact, the tip-height is reduced until the contact has been achieved. At that distance, the two potential wells overlap and appear as one well (Fig. 8b). The atom/molecule is then easily transferred to the tip.

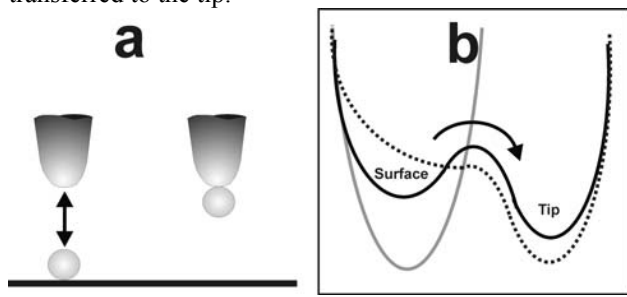

**Fig. 8.** Vertical manipulation (VM). (a) A schematic drawing shows the process. (b) The double-potential well model. The black (solid), dash and gray curves represent the shape of potentials (the well shapes are working assumptions) at an image-height, under an electric field, and at the tip-atom/molecule contact, respectively.

In the case of vertical manipulation of CO molecules, a temporary tunneling electron attachment into a $2\pi^*$ anti-bonding state of CO leads to the breaking of the CO-Cu bond first, and the resultant excited CO molecule can either jump to a nearby Cu surface site or to the tip-apex[4, 6]. The CO can be transferred back to the surface using the same process. One useful application of VM is to modify the STM-tip. A single atom/molecule tip can be fabricated by deliberately transferring an atom/molecule to the tip apex[6, 11, 23, 31, 39, 77]. This improves the tip sharpness and thus the STM image resolution can be enhanced[11, 23, 39, 77]. The chemical identity of the tip-apex can also be known from this process. The molecule-tips are useful in molecular recognition imaging. For example, CO and oxygen can be distinguished when a CO-tip is used to image[6]. An application is demonstrated in Fig. 9. The VM procedure can also be used to transport the atoms/molecules across a surface[6, 31, 32, 39, 57]. In this case the VM process is analogous to loading and unloading operation of a crane at a construction site in our macroscopic world.

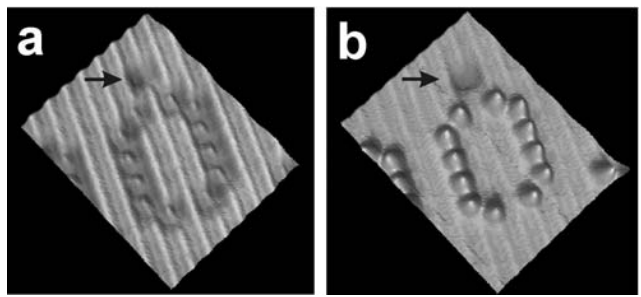

**Fig. 9** Imaging with a CO-tip. (a) The CO molecules that form '0' on a Cu(211) surface appear as depressions when imaged with a normal tip. (b) The CO molecules appear as protrusions when imaged with a CO-tip. An unknown adsorbate (shown with arrow) is still remains as a depression in both images, thus chemical contrast is achieved [Courtesy of J. Repp, IBM Zurich].

## VI. INELASTIC TUNNELING INDUCED MANIPULATION

Controlled excitations of atoms and molecules can be performed by using inelastic-electron-tunneling (IET) induced manipulation processes[38-40, 49, 55, 56, 69, 74, 80, 83] (Fig. 10). In an IET manipulation, low energy tunneling electrons or holes (if electron tunneling is from the surface to the tip) are injected to the atom/molecule located on a surface by positioning the tip above the target. The tunneling electron energy is transferred to an atom/molecule through a resonance state leading to various excitations[25]. In an IET manipulation process, the maximum tunneling-electron-energy can be controlled by adjusting the applied bias. The excitation rate can be varied by changing the tunneling current. If the tunneling current changes, the number of tunneling electrons passing through the atom/molecule will change accordingly and hence, the probability and the rate of excitation will also vary. Two

IET processes, single and multiple excitations, can be distinguished[83]. An excitation event caused by a single electron energy transfer is known as the single excitation[40, 83]. Several electrons are involved in a multiple excitation process, and it can be explained by using a harmonic oscillator model (Fig. 10b). Here, the electron-energy transfer causes the atom/molecule to excite at an energy level, and subsequent energy transfer by other electrons raises the excited energy level further. The multiple excitation process require a longer life time of the excited state in order to couple the excitations. In case of molecular excitations; rotational, vibrational and electronic excitation can be induced[55, 83]. Bond breaking and bond formation between the molecular fragments can also be realized by using IET manipulation.

It is possible to determine the number of tunneling electrons involved in an IET process, by using the relation[40, 83];

$$R \propto I^N$$

where "R" is the excitation rate, "I" is the tunneling current, and "N" is the number of tunneling electrons. In case of oxygen dissociation on Pt(111), a pioneering work of IET dissociation done in the group of Ho[83], both single and multiple excitation processes have been successfully demonstrated. Controlled dissociation of polyatomic molecules using tunneling electrons is more complex than that of diatomic species like $O_2$. This is because polyatomic molecules can have a variety of bonds and hence, the tunneling process may involve more than one bond. As an illustration, the dissociation of two C-I bonds of a $C_6H_4I_2$ molecule is shown in Fig 11.

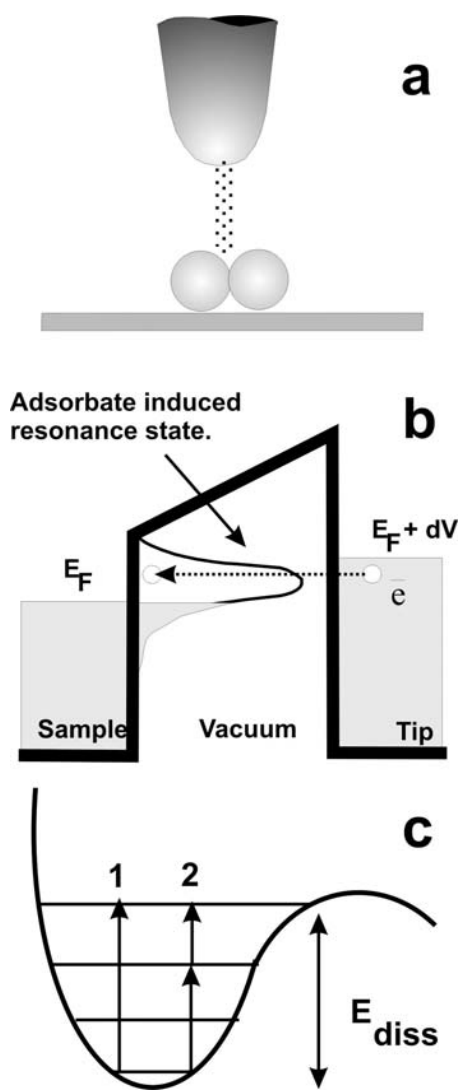

**Fig. 10** IET manipulation. (a) A schematic drawing shows an IET dissociation. (b) Inelastic tunneling electrons are injected to the molecule through an adsorbate-induced resonance state. (c) Single and multiple excitation processes in case of IET induced bond breaking.

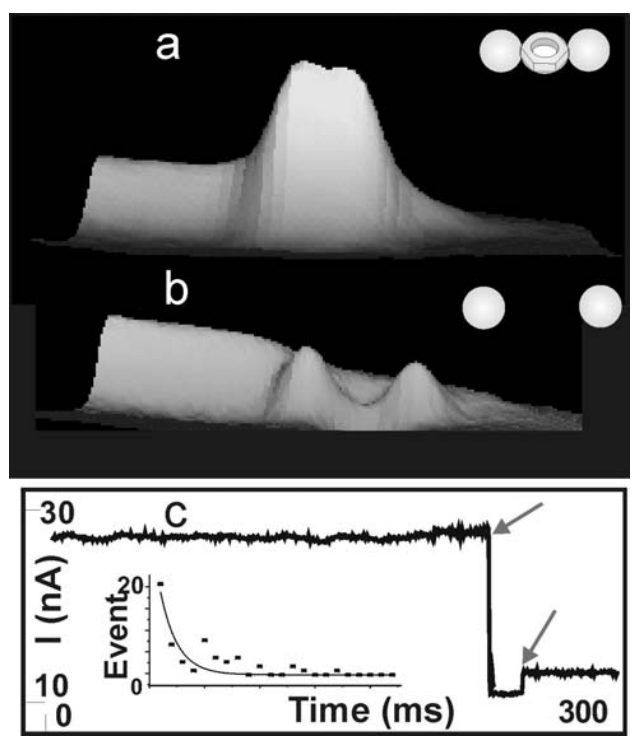

**Fig. 11.** Iodine abstraction. (a) An adsorbed p-diiodobenzene ($C_6H_4I_2$) molecule at Cu(111) a screw dislocation. (b) After breaking the two C-I bonds of the molecule using IET process, the two iodine atoms remain adsorbed. (c) The changes in the current (shown with arrows) are due to bond-breaking events. Inset shows the C-I bond dissociation probability collected from 70 dissociations.

### VII. ELECTRIC FIELD INDUCED MANIPULATION

The electric field induced manipulation uses the voltage applied between the tip and sample to manipulate single atoms/molecules. By changing the bias polarity, the atom/molecule having a dipole can experience either attractive or repulsive force from the tip. For example, in case of a vertical manipulation of Xe atoms[21] or CO molecules[4, 6], the changing of the electric field polarity can lead to direct desorption and readsorption of the atom/molecule between the tip and the surface. Smaller molecules, like CO, can be moved laterally across a surface using this manipulation technique[49]. When a high electric field (> 3V) is used, bond breaking of molecules can be performed[1, 2, 20, 62, 75]. The group of Avouris has first demonstrated dissociation of $B_{10}H_{14}$ and $O_2$ molecules on Si(111) using high STM biases (≥ 4V and ≥6V respectively)[20]. At the field emission regime where the bias is higher than the work function of the tip, the tip acts as an electron emission gun. In this case, the number of electrons passing through an atom/molecule is no longer controllable.

## VIII. COMBINED USAGE OF STM MANIPULATIONS

Innovative experimental schemes to investigate specific properties of single atoms/molecules can be developed by combining a variety of STM manipulation procedures[31, 33, 35, 39, 40]. A good example is the induction of single molecule Ullmann reaction[40] on a Cu(111) surface where all the basic reactions steps; dissociation, diffusion and association, have been realized by using several manipulation techniques in a step-by-step manner. This chemical reaction sequence results in synthesis of a biphenyl molecule out of two iodobenzeze. To provide the readers a taste of a combined manipulation experiment, a molecular *shooting* scheme recently developed by the author[35] is presented. Here, physisorbed sexiphenyl molecules on Ag(111) are used as bullets and they are shot toward targets by manipulating with the STM-tip.

Sexiphenyl is composed of six π-rings, which are connected like a linear chain[11, 35]. Sexiphenyl is weakly bound to Ag(111) surface, and this makes it difficult to laterally manipulate the molecule with an atomic precision. During the LM process, the molecule occasionally slides further across the surface after retracting the tip. Using this property of molecule on Ag(111), a molecular shooting scheme has been developed[35]. To shoot the molecule, it is dragged with the tip towards the long molecular-axis direction for a few nm. When the molecule is released by retracting the tip, it continues to propagate up to 30 nm distance on Ag(111) surface at 6K. During this shooting process, the molecular trajectories are not in straight lines because the random electron-standing-waves and defects on the surface can cause the molecule to deflect from its original propagation path.

In order to achieve atomically-straight trajectory of the molecules and to probe the mechanism of molecular propagation, a uniform electronic structural environment is necessary. For this, an electron resonator having linear parallel standing-wave front is constructed (Fig 12). The silver atoms used for the construction are extracted from the native substrate by means of a tip-crash procedure described above[36]. Shooting sexiphenyl with the STM-tip inside this quantum structure provide atomically straight trajectories. To demonstrate the atomic scale control, two sexiphenyl molecules (bullets) and two silver atoms (targets) are positioned at opposite ends of the linear resonator (Fig. 12a). Then one of the molecules is shot toward the target silver atom. As soon as the molecule hits the target, a sexiphenyl-Ag complex is formed (Fig. 12b). This shooting can be performed both at the standing wave minimum and maximum tracks. Thus, uniform electronic structural environment is the key for an atomically controlled shooting process.

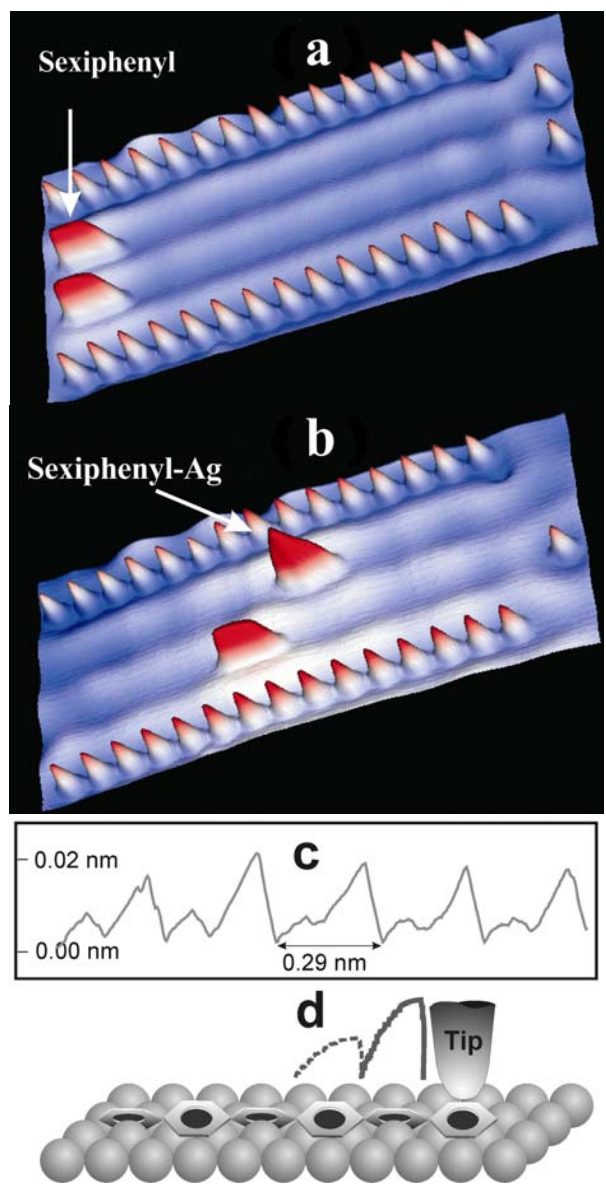

**Fig. 12.** Molecular shooting: (a) Two sexiphenyls (left) and two target atoms (right) are positioned along the standing wave track. A silver-sexiphenyl complex is formed by shooting the upper sexiphenyl. (b) The Ag-sexiphenyl complex (upper) and the bare sexiphenyl (lower) are repositioned for a visual comparison. (c) A periodic low-high peak manipulation signal is observed during LM. (d) A schematic drawing illustrates alternating twist of sexiphenyl π-rings and the tip position during manipulation. [Imaging parameters: $V_t$ = 30 mV, $I_t$= 1.1 nA, 16 x 26 $nm^2$].

It is known that the π-rings of sexiphenyl are alternately twisted in the gas phase with a torsional angle between 20 to 40 degrees due to the steric repulsion. Sexiphenyl on Ag(111) has an alternately twisted π-ring configuration[11]. To determine the π-ring twisting angle and internal conformation changes during its movement, a controlled LM process is performed inside the electron resonator. The tip is positioned at an edge of a sexiphenyl π-ring during LM to detect the π-ring movement. The manipulation signal reveals periodic low-high tip-height peaks with hcp-fcc lattice distance of Ag(111) (Fig. 12c). Because of the constant current scanning mode, the tip needs to retract in order to maintain the current constant when the π-ring edge is tilted upward. Thus, the higher peak corresponds to the upward tilted π-ring edge configuration. When the π-ring edge is tilted downward, a low-peak signal is produced. This reveals that the sexiphenyl moves across the surface by flipping its alternate π-ring twist to and fro, like a *caterpillar* moving on the ground. The molecule shooting mechanism is triggered by the mechanical energy stored in its twisted π-ring configuration during dragging with the STM-tip[35]. From the manipulation geometry, the torsional angle of sexiphenyl on Ag(111) can be determined as 11.4 degree[11]. This combined manipulation experiment set an example where the experiments can be conducted within a few tens of nm square area and the single molecule level information, which can not be accessed by other experimental techniques, can be obtained.

## IX. OUTLOOK

The progress in manipulation techniques has not only extended the application of STM but also created a new field of nanoscale engineering. Construction of atomic scale structures, such as quantum corrals and electron resonators, allows the study of quantum phenomena locally[12, 17, 33, 35, 37, 61]. Inducing chemical reactions at single molecule level with the STM-tip gives a distinctive insight to the reaction mechanism at a fundamental level[31, 40, 47]. Chemical reactions like the Ullmann reaction can be verified on an entirely new platform[39, 40]. Novel chemical reaction pathways can be discovered[31]. Such manipulation experiments are fascinating and, without a doubt, greatly contribute to the advancement of science. Most STM atom/molecule manipulations require sophisticated instruments to achieve a precise control at the tip-sample junction. Thus only a few research groups in the world have been able to use such manipulation techniques as compared to a large number of research groups using STM imaging and spectroscopy methods. However, this trend is gradually changing in recent years. Room temperature manipulations have been able to perform on large molecules[16, 28, 44, 81, 82]. The atom manipulation process can also be automated, which increases the speed of atomistic construction substantially. With all the exciting advances in manipulation techniques, the STM atom/molecule manipulation will continue to be a powerful analytical technique of nanoscience, just as the STM imaging and spectroscopy. STM imaging has greatly contributed to surface science and semiconductor industry. It is certain that the STM manipulation will follow a similar trend coupled with equally outstanding contributions to nanoscience research and future nanotech industry.

**Acknowledgement:** The support provided by the US Department of Energy grant DE-FG02-02ER46012, the NSF- DMR 0304314, and the Ohio University CMSS Program are gratefully acknowledged.

**LITERATURE CITED**
[1] Ph. Avouris, R.E. Walkup, A.R. Rossi, T.-C. Shen, G.C. Abeln, J.R. Tucker, and J.W. Lyding, "STM-induced H atom desorption from Si(100): Isotope effects and site selectivity", Chem. Phys. Lett. **257**, 148 (1996).
[2] Ph. Avouris, R.E. Walkup, A.R. Rossi, H.C. Akpati, P. Nordlander, T.-C. Shen, J.W. Lyding, and G.C. Abeln, "Breaking individual chemical bonds via STM-induced excitations", Surf. Sci. **363**, 368 (1996).
[3] L. Bartels, B.V. Rao, and A. Liu, "Translation and rotation of a haloaromatic thiol", Chem. Phys. Lett. **385**, 36 (2004).
[4] L. Bartels, G. Meyer, K.-H. Rieder, D. Velic, E. Knoesel, A. Hotzel, M. Wolf, and G. Ertl, "Dynamics of electron-induced manipulation of individual CO molecules on Cu(111)", Phys. Rev. Lett. **80**, 2004 (1998).
[5] L. Bartels, G. Meyer, and K.-H. Rieder, "Basic steps of lateral manipulation of single atoms and diatomic clusters with a scanning tunneling microscope tip", Phys. Rev. Lett. **79**, 697 (1997).
[6] L. Bartels, G. Meyer, and K.-H. Rieder, "Controlled vertical manipulation of single CO molecules with the scanning tunneling microscope: A route to chemical contrast", Appl. Phys. Lett. **71**, 213 (1997).
[7] Besocke, K., "An easily operatable scanning tunneling microscope", Surf. Sci. **181**, 145 (1987).
[8] P.H. Beton, A.W. Dunn, and P. Moriarty, "Manipulation of C60 molecules on a Si surface", Appl. Phys. Lett. **67**, 1075 (1995).
[9] G. Binnig, H. Rohrer, C. Gerber, and E. Weibel, "7X7 Reconstruction on Si(111) resolved in real space", Phys. Rev. Lett. **50**, 120 (1983).
[10] X. Bouju, C. Joachim, and C. Girard, "Single-atom motion during a lateral STM manipulation" Phys. Rev. B **59**, R7845 (1999).
[11] K.-F. Braun, and S.-W. Hla, "Probing the conformation of physisorbed molecules at the atomic-scale using STM manipulation", Nano Lett. **5,** 73 (2005).
[12] K.F. Braun, and K.-H. Rieder, "Engineering electronic lifetimes in artificial atomic structures", Phys. Rev. Lett. **88**, 096801 (2002).
[13] C.J. Chen. Introduction to Scanning Tunneling Microscopy. Oxford University Press. 1993.
[14] S.G. Corcoran, R.J. Colton, E.T. Lilleodden, and W.W. Gerberich, "Anomalous plastic deformation at surfaces: Nanoindentation of gold single crystals", Phys. Rev. B **55**, R16057 (1997).
[15] P.Z. Coura, S.B. Legoas, A.S. Moreira, F. Sato, V. Rodrigues, S.O. Dantas, D. Ugarte, and D.S. Galväo, "On the structural and stability features of linear atomic suspended chains formed from gold nanowires stretching", Nano. Lett. **4**, 1187 (2004).
[16] M.T. Cuberes, R.R. Schlittler, and J.K. Gimzewski, "Room-temperature repositioning of individual C60 molecules at Cu steps: Operation of a molecular counting device", Appl. Phys. Lett. **69**, 3016 (1996).

[17] M.F. Crommie, C.P. Lutz, and D.M. Eigler, "Confinement of electrons to quantum corrals on a metal-surface", Science **262**, 218 (1993).

[18] G. Dujardin, A.J. Mayne, and F. Rose, "Temperature control of electronic channels through a single atom", Phys. Rev. Lett. **89**, 036802 (2002).

[19] G. Dujardin, A. Mayne, O. Robert, F. Rose, C. Joachim, and H. Tang, "Vertical manipulation of individual atoms by a direct STM tip-surface contact on Ge(111)", Phys. Rev. Lett. **80**, 3085 (1998).

[20] G. Dujardin, R.E. Walkup, and Ph. Avouris, "Dissociation of individual molecules with electrons from the tip of a scanning tunneling microscope", Science **255**, 1232 (1992).

[21] D.M. Eigler, C.P. Lutz, and W.E. Rudge, "An atomic switch realized with the scanning tunnelling microscope", Nature **352**, 600 (1991).

[22] D.M. Eigler, and E.K. Schweizer, "Positioning single atoms with a scanning tunneling microscope", Nature **344**, 524 (1990).

[23] S. Folsch, P. Hyldgaard, R. Koch, and K.H. Ploog, "Quantum confinement in monatomic Cu chains on Cu(111)", Phys. Rev. Lett. **92**, 056803 (2004).

[24] O.R. de la Fuente, J.A. Zimmerman, M.A. González, J. de la Figuera, J.C. Hamilton, W.W. Pai, and J.M. Rojo, "Dislocation emission around nanoindentations on a (001) fcc metal surface studied by scanning tunneling microscopy and atomistic simulations", Phys. Rev. Lett. **88**, 036101 (2002).

[25] J.W. Gadzuk, "Resonance-assisted, hot-electron-induced desorption", Surf. Sci. **342**, 345 (1995).

[26] A. Gannepalli, and S.K. Mallapragada, "Atomistic studies of defect nucleation during nanoindentation of Au(001)", Phys. Rev. B **66**, 104103 (2002).

[27] J. Gaudioso, and W. Ho, "Single-molecule vibrations, conformational changes, and electronic conductivity of five-membered heterocycles", J. Am. Chem. Soc. **123**, 10095 (2001).

[28] S.J.H. Griessl, M. Lackinger, F. Jamitzky, T. Markert, M. Hietschold, and W.M. Heckl, "Room-temperature scanning tunneling microscopy manipulation of single C-60 molecules at the liquid-solid interface: Playing nanosoccer", J. Phys. Chem. B **108**, 11556 (2004).

[29] M. Grobis, K. H. Khoo, R. Yamachika, Xinghua Lu, K. Nagaoka, Steven G. Louie, M. F. Crommie, H. Kato, and H. Shinohara, "Spatially dependent inelastic tunneling in a single metallofullerene", Phys. Rev. Lett. **94**, 13602 (2004).

[30] L. Gross, F. Moresco, M. Alemani, H. Tang, A. Gourdon, C. Joachim, and K.-H. Rieder, "Lander on Cu(211) - selective adsorption and surface restructuring by a molecular wire", Chem. Phys. Lett. **371**, 750 (2003).

[31] J. R. Hahn, and W. Ho, "Oxidation of a single carbon monoxide molecule manipulated and induced with a scanning tunneling microscope", Phys. Rev. Lett. **87**, 166102 (2001).

[32] J.A. Heinrich, Gupta, J.A., C.P. Lutz, J.A., and D.M. Eigler, "Single-atom spin-flip spectroscopy", Science **306**, 466 (2004).

[33] J.A. Heinrich, C.P. Lutz, J.A. Gupta, J.A., and D.M. Eigler, "Molecule cascade", Science **298**, 1381 (2002).

[34] E.J. Heller, M.F. Crommie, C.P. Lutz, and D.M. Eigler, "Scattering and absorption of surface electron waves in quantum corrals", Nature **369**, 464 (1994).

[35] S.-W. Hla, K.-F. Braun, B. Wassermann, and K.-H. Rieder, "Controlled low-temperature molecular manipulation of sexiphenyl molecules on Ag(111) using scanning tunneling microscopy", Phys. Rev. Lett. **93**, 208302 (2004).

[36] S.-W. Hla, K.-F. Braun, V. Iancu, and A. Deshpande, "Single atom extraction by scanning tunneling microscope tip-crash and nanoscale surface engineering", Nano Lett. **4**, 1997 (2004).

[37] S.W. Hla, K. F. Braun, and K.-H. Rieder, "Detailed atom movement mechanisms during a quantum corral construction", Phys. Rev. B **67,** 201402R (2003).

[38] S.W. Hla, G. Meyer, and K.-H. Rieder, "Selective bond breaking of single iodobenzene molecules with a scanning tunneling microscope tip", Chem. Phys. Lett **370**, 431 (2003).

[39] S.-W. Hla, G. Meyer, and K.-H. Rieder, "Inducing single-molecule chemical reactions with a UHV-STM: A new dimension for nanoscience and technology", ChemPhysChem **2**, 361 (2001).

[40] S.W. Hla, L. Bartels, G. Meyer, and K.-H. Rieder, "Inducing all steps of a chemical reaction with the scanning tunneling microscope tip: Towards single molecule engineering", Phys. Rev. Lett. **85**, 2777 (2000).

[41] S.-W. Hla, A. Kühnle, L. Bartels, G. Meyer, and K.-H. Rieder, "Controlled lateral manipulation of single diiodobenzene molecules on the Cu(111) surface with the tip of a scanning tunneling microscope", Surf. Sci. **454-456**, 1079 (2000).

[42] T. Jamneala, V. Madhavan, and M.F. Crommie, "Kondo response of a single antiferromagnetic chromium trimer", Phys. Rev. Lett. **87**, 256804 (2001).

[43] C. Joachim, H. Tang, F. Moresco, G. Rapenne, and G. Meyer, "The design of a nanoscale molecular barrow", Nanotechnology **13**, 330 (2002).

[44] T.A. Jung, R.R. Schlittler, J.K. Gimzewski, H. Tang, and C. Joachim, "Controlled roomtemperature positioning of individual molecules: Molecular flexure and motion", Science **271**, 181 (1996).

[45] M. Kageshima, H. Ogiso, and H. Tokumoto, "Lateral forces during manipulation of a single C-60 molecule on the Si(001)-2 x 1 surface", Surf. Sci. **517**, L557 (2002).

[46] J.D. Kiely, and J.E. Houston, "Nanomechanical properties of Au (111), (001), and (110) surfaces", Phys. Rev. B **57**, 12588 (1998).

[47] Y. Kim, T. Komeda, and M. Kawai, "Single-molecule reaction and characterization by vibrational excitation", Phys. Rev. Lett. **89**, 126104 (2002).

[48] T. Komeda, Y. Kim, Y. Fujita, Y. Sainoo, and M. Kawai, "Local chemical reaction of benzene on Cu(110) via STM-induced excitation", J. Chem. Phys. **120**, 5347 (2004).

[49] T. Komeda, Y. Kim, M. Kawai, B.N.J. Persson, and H. Ueba, "Lateral hopping of molecules induced by excitation of internal vibration mode", Science **295**, 2055 (2002).

[50] Y. Kondo, and K. Takayanagi, "Synthesis and characterization of helical multi-shell gold nanowires", Science **289**, 606 (2000).

[51] A. Kühnle, G. Meyer, S. W. Hla, and K. -H. Rieder, "Understanding atom movement during lateral manipulation with the STM tip using a simple simulation method", Surf. Sci. **499**, 15 (2002).

[52] O. Kurnosikov, J.T. Kohlhepp, and W.J.M. de Jonge, "Can surface embedded atoms be moved with an STM tip?", Eur. Phys. Lett. **64**, 77 (2003).

[53] J. Lagoute, K. Kanisawa, and S. Folsch, "Manipulation and adsorption-site mapping of single pentacene molecules on Cu(111)", Phys. Rev. B. **70**, 245415 (2004).

[54] U. Landman, W.D. Luedtke, N.A. Burnham, and R.J. Colton, "Atomistic mechanisms and dynamics of adhesion, nanoidentation and fracture", Science **248**, 454 (1990).

[55] L.J. Lauhon, and W. Ho, "Control and characterization of a multistep unimolecular reaction" Phys. Rev. Lett. **84**, 1527 (2000).

[56] L.J. Lauhon, and W. Ho, "Single-molecule chemistry and vibrational spectroscopy: Pyridine and benzene on Cu(001)", J. Phys. Chem. **104**, 2463 (2000).

[57] H.J. Lee, and W. Ho, "Single-bond formation and characterization with a scanning tunneling microscope", Science **286**, 1719 (1999).

[58] J.T. Li, W.-D. Schneider, and R. Berndt, "Low-temperature manipulation of Ag atoms and clusters on a Ag(110) surface", Appl. Phys. A. **66**, S675 (1998).

[59] R. Lin, K.-F. Braun, H. Tang, U.J. Quaade, F.C. Krebs, G. Meyer, C. Joachim, K.-H. Rieder, and K. Stokbro, "Imaging and manipulation of a polar molecule oil Ag(111)", Surf. Sci. **477**, 198 (2001).

[60] I.-W. Lyo, and Ph. Avouris, "Field-induced nanometer- to atomic-scale manipulation of silicon surfaces with the STM", Science **253**,173 (1991).

[61] H.C. Manoharan, C.P. Lutz, and D.M. Eigler, "Quantum mirages formed by coherent projection of electronic structure", Nature **403**, 512 (2000).

[62] R. Martel, Ph. Avouris, and I.-W. Lyo, "Molecularly adsorbed oxygen species on Si(111)-(7*7): STM-induced dissociative attachment studies", Science **272**, 385 (1996).

[63] G. Meyer, S. Zöphel, and K.-H. Rieder, "Controlled manipulation of ethen molecules and lead atoms on Cu(211) with a low temperature scanning tunneling microscope", Appl. Phys. Lett. **69**, 3185 (1996).

[64] G. Meyer, S. Zöphel, and K.-H. Rieder, "Scanning tunneling microscopy manipulation of native substrate atoms: A new way to obtain

[65] G. Meyer, S. Zöphel and K.-H. Rieder, "Manipulation of atoms and molecules with a low temperature scanning tunneling microscope", Appl. Phys. A **63**, 557 (1996).

[66] Meyer, G., "A simple low-temperature ultrahigh-vacuum scanning tunneling microscope capable of atomic manipulation", Rev. Sci. Instrum. **67**, 2960 (1996).

[67] G. Meyer, B. Neu, and K.-H. Rieder, "Controlled lateral manipulation of single molecules with the scanning tunneling microscope", Appl. Phys. A. **60**, 343 (1995).

[68] F. Moresco, G. Meyer, K.-H. Rieder, J. Ping, H. Tang, and C. Joachim, "TBPP Molecules on copper surfaces: a low temperature scanning tunneling microscope investigation", Surf. Sci. **499**, 94 (2002).

[69] F. Moresco, G. Meyer, K.-H. Rieder, H. Tang, and C. Joachim, "Conformational changes of single molecules induced by scanning tunneling microscopy manipulation: A route to molecular switching", Phys. Rev. Lett. **86**, 672 (2001).

[70] F. Moresco, G. Meyer, K.-H. Rieder, H. Tang, A. Gourdon, and C. Joachim, "Recording intramolecular mechanics during the manipulation of a large molecule", Phys. Rev. Lett. **87**, 088302 (2001).

[71] F. Moresco, G. Meyer, K.-H. Rieder, H. Tang, A. Gourdon, and C. Joachim, "Low temperature manipulation of big molecules in constant height mode", Appl. Phys. Lett. **78**, 306 (2001).

[72] G. V. Nazin, X. H. Qiu, and W. Ho, "Visualization and spectroscopy of a metal-molecule-metal bridge", Science **302**, 77 (2003).

[73] F.E. Olsson, M. Persson, A.G. Borisov, J.P. Gauyacq, J. Lagoute, and S. Folsch, "Localization of the Cu(111) surface state by single Cu adatoms", Phys. Rev. Lett. **93**, 206803 (2004).

[74] J.I. Pascual, N. Lorente, Z. Song, H. Conrad, and H.P. Rust, "Selectivity in vibrationally mediated single-molecule chemistry", Nature **423**, 525 (2003).

[75] N.J. Persson, Ph. Avouris, "Local bond breaking via STM-induced excitations: The role of temperature", Surf. Sci. **390**, 45 (1997).

[76] X.H. Qiu, G.V. Nazin, and W. Ho, "Vibronic states in single molecule electron transport", Phys. Rev. Lett. **92**, 206102 (2004).

[77] J. Repp, G. Meyer, S.M. Stojkovic, A. Gourdon, and C. Joachim, "Molecules on insulating films: Scanning-tunneling microscopy imaging of individual molecular orbitals", Phys. Rev. Lett. **94**, 026803 (2005).

[78] J. Repp, G. Meyer, F.E. Olsson, M. Persson, "Controlling the charge state of individual gold adatoms", Science **305**, 493 (2004).

[79] J.J. Saenz, and N. Garcia, "Quantum atom switch: tunneling of Xe atoms", Phys. Rev. B **47**, 7537 (1993).

[80] Y. Sainoo, Y. Kim, T. Komeda, and M. Kawai, "Inelastic tunneling spectroscopy using scanning tunneling microscopy on trans-2-butene molecule: Spectroscopy and mapping of vibrational feature", J. Chem. Phys. **120**, 7249 (2004).

[81] P.A. Sloan, and R.E. Palmer, "Two-electron dissociation of single molecules by atomic manipulation at room temperature", Nature **434**, 367 (2005).

[82] P.A. Sloan, M.F.G. Hedouin, R.E. Palmer, and M. Persson, "Mechanisms of molecular manipulation with the scanning tunneling microscope at room temperature: Chlorobenzene/Si(111)-(7x7) ", Phys. Rev. Lett. **91**, 118301 (2003).

[83] B. C. Stipe, M.A. Rezaei, W. Ho, S. Gao, M. Persson, and B.I. Lundqvist, "Single-molecule dissociation by tunneling electrons", Phys. Rev. Lett. **78**, 4410 (1997).

[84] J. A. Stroscio, and R. J. Celotta, "Controlling the dynamics of a single atom in lateral atom manipulation", Science **306**, 242 (2004).

[85] J.A. Stroscio, and D.M. Eigler, "Atomic and molecular manipulation with the scanning tunneling microscope", Science **254**, 1319 (1991).

[86] H. Tang, M.T. Cuberes, C. Joachim, and J.K. Gimzewski, "Fundamental considerations in the manipulation of a single C60 molecule on a surface with an STM", Surf. Sci. **386**, 115 (1997).

[87] I.S. Tilinin, M.A. Van Hove, and M. Salmeron, "Tip-surface transfer of adatoms in AFM/STM: effect of quantum oscillations", Appl. Surf. Sci. **130-132**, 676 (1998).

[88] E. Tosatti, and S. Prestipino, "Weird gold nanowires", Science **289**, 561 (2000).

[89] T. T. Tsong, "Effects of an electric field in atomic manipulations", Phys. Rev. B **44**, 13703 (1991).

[90] P. Vettiger, M. Despont, U. Drechsler, U. Durig, W. Haberle, M.I. Lutwyche, H.E Rothuizen, R. Stutz, R. Widmer, and G.K. Binnig, "The "Millipede" - More than one thousand tips for future AFM data storage" IBM J. Res. Dev. **44**, 323 (2000).

12